\def\year{2019}\relax
\documentclass[letterpaper]{article} 
\usepackage{aaai19}  
\usepackage{times}  
\usepackage{helvet}  
\usepackage{courier}  
\usepackage{url}  
\usepackage{graphicx}  

\usepackage{xcolor}
\usepackage{dblfloatfix}
\usepackage{makecell} 

\usepackage{amsfonts} 
\usepackage{amsmath} 

\title{Automatic Detection and Compression for Passive \\ Acoustic Monitoring of the African Forest Elephant}

\author{Johan Bjorck \textsuperscript{1}
Brendan H. Rappazzo\textsuperscript{1}
Di Chen\textsuperscript{1} \\
{\Large\bf Richard Bernstein\textsuperscript{1}
Peter H. Wrege\textsuperscript{2}
Carla P. Gomes\textsuperscript{1}}  \\
\textsuperscript{1}{Dept. of Computer Science, Cornell University, Ithaca, NY}\\
\textsuperscript{2}{Bioacoustics Research Program, Cornell Lab of Ornithology, Ithaca, NY}}

\begin{document}
\maketitle
\begin{abstract}
In this work, we consider applying machine learning to the analysis and compression of audio signals in the context of monitoring elephants in sub-Saharan Africa. Earth’s biodiversity is increasingly under threat by sources of anthropogenic change (e.g. resource extraction, land use change, and climate change) and surveying animal populations is critical for developing conservation strategies. However, manually monitoring tropical forests or deep oceans is intractable. For species that communicate acoustically, researchers have argued for placing audio recorders in the habitats as a cost-effective and non-invasive method, a strategy known as passive acoustic monitoring (PAM). In collaboration with conservation efforts, we construct a large labeled dataset of passive acoustic recordings of the African Forest Elephant via crowdsourcing, compromising thousands of hours of recordings in the wild. Using state-of-the-art techniques in artificial intelligence we improve upon previously proposed methods for passive acoustic monitoring for classification and segmentation. In real-time detection of elephant calls, network bandwidth quickly becomes a bottleneck and efficient ways to compress the data are needed. Most audio compression schemes are aimed at human listeners and are unsuitable for low-frequency elephant calls. To remedy this, we provide a novel end-to-end differentiable method for compression of audio signals that can be adapted to acoustic monitoring of any species and dramatically improves over näive coding strategies.
\end{abstract}

\section{Introduction}

Poaching, illegal logging, and infrastructure expansions are some of many current threats to biodiversity, and large mammals are particularly susceptible. To effectively allocate conservation resources and develop conservation strategies, endangered animal populations need to be accurately and economically surveyed, but for species that roam large or inaccessible areas monitoring by humans becomes intractable. A promising approach for species communicating via acoustic signals is passive acoustic monitoring (PAM), which involves the use of autonomous recording devices scattered throughout habitats that record animal vocalizations. Compared to video monitoring, acoustic monitoring is not limited by line of sight, is typically considerably cheaper and requires less bandwidth for transferring the data. However, extracting useful data from these soundscapes is non-trivial and automatic approaches are necessary.

\begin{figure}
\centering
\includegraphics[width=1.0\columnwidth]{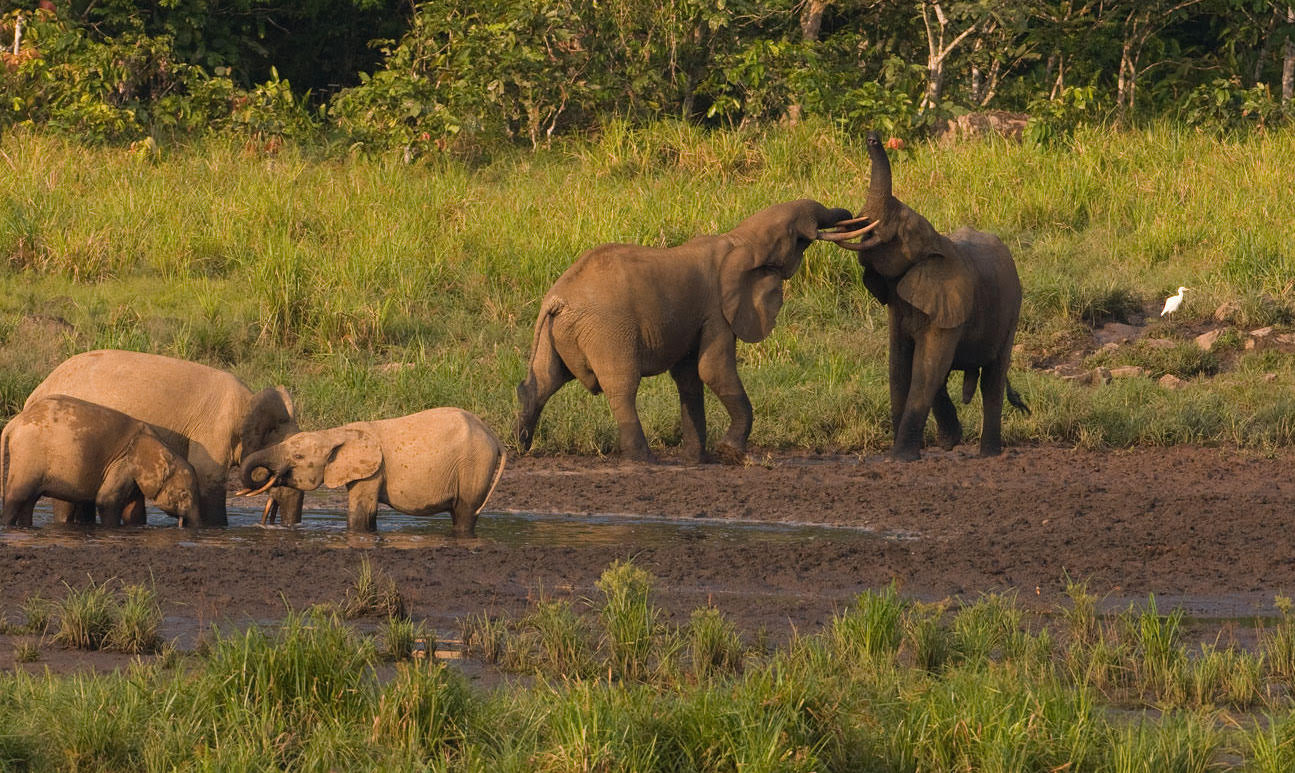}
\caption{The African forest elephant (Loxodonta cyclotis) is the smallest of the three extant elephant species, a keystone species in the rainforests of the Congo Basin, and is entirely relied upon by many trees to disperse their seeds \cite{campos2011megagardeners}. Due to their highly-valued ivory tusks, the elephant is a typical target for poachers in central Africa and the population has fallen by more than $60\%$ in the last decade \cite{bbc}. Population monitoring is critical for the elephant's survival, and in this work, we consider combining passive acoustic monitoring and artificial intelligence towards this end.}
\end{figure}

\begin{figure*}[b!]
\centering
\includegraphics[width=1.9\columnwidth]{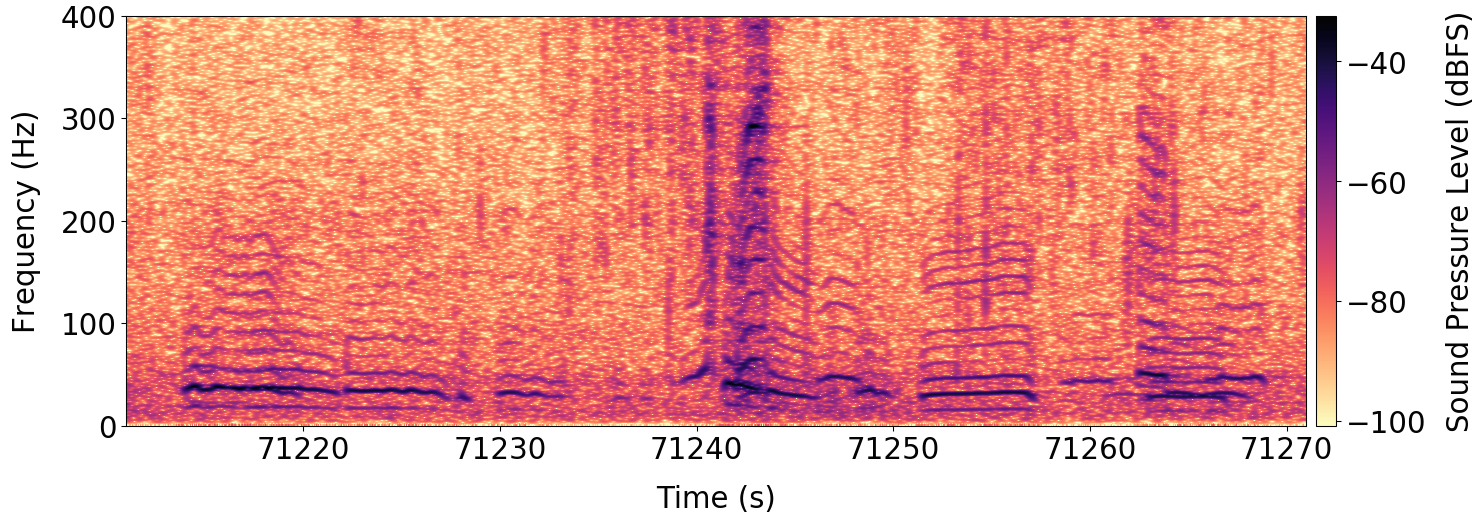}
\caption{A spectrogram of several elephant rumble vocalizations within a 60-second segment of sound. The rich harmonic structure is typical of rumbles, however, since higher frequency elements attenuate rapidly with distance, recording these higher frequency elements depends on source amplitude and distance. Thus, it is difficult to infer distance from harmonic structure alone.}
\label{fig:typical_calls}
\end{figure*}

In this work, we consider PAM in the context of monitoring the African forest elephant. To enable real-time acoustic monitoring one needs to quickly and accurately detect elephants and potential threats to them -- a classical challenge of classification and segmentation. Leveraging recent advances in neural networks, we improve upon previous methods in automating PAM. In real-time threat-detection and population monitoring the bandwidth of the wireless networks becomes a bottleneck, and one additionally has to use efficient data representations to only communicate the necessary information. In many lossy compression schemes, signal components inaudible to humans such as low frequencies are given low bit-rates, which in the context of low-frequency elephant calls is a poor strategy. Using a differentiable proxy for non-differentiable bit truncation, we are able to cast this problem as an end-to-end differentiable setup, which can be trained via stochastic gradient descent (SGD) to get improved compression.

We focus on the African Forest Elephant both due to its biological importance and the loud calls by which it communicates. This elephant is a keystone species in the rainforests of the Congo Basin, the second largest expanse of rainforest on earth and among the most speciose. Conserving viable populations of forest elephants protects local biodiversity, but the expansiveness of the rainforest and the difficulty of monitoring animals within it makes manual monitoring problematic. Since elephants communicate over long distances via infrasonic signals referred to as rumbles \cite{hedwig2018not} they are particularly suited to an acoustic approach. These characteristic vocalizations provide information on occupancy, landscape use, population size, and the effects of anthropogenic disturbances \cite{wrege2017acoustic}.

The contributions of our work are to \textbf{1)} construct a large dataset of real-world elephant vocalizations from central Africa via passive acoustic monitoring, \textbf{2)} surpass previously proposed methods for automatic PAM in the context of elephant vocalizations via state-of-the-art artificial intelligence techniques, and \textbf{3)} introduce a novel end-to-end differentiable technique for audio-compression, that can balance the bit-rates between different frequency channels to the unique characteristics of the monitored species.
 
\section{The Dataset} \label{sec:data}

\subsection{Data collection}

Established in 2000, the Elephant Listening Project (ELP) uses acoustic methods to study the ecology and behavior of forest elephants in order to improve evidence-based decision making concerning their conservation. ELP has recorded sounds from over 150 different locations, amassing more than 700,000 hours of recordings. These varying environments provide the source material for generating training data for algorithm development. The dataset we consider in this work was collected between 2007 and 2012 from three sites in Gabon and one in the Central African Republic, which will be referred to as Ceb1, Ceb4, Dzanga, and Jobo. A map showing these locations is given in Figure \ref{fig:map} of the Appendix. At all locations, a single recording device was placed in a tree 7-10 meters above the ground near forest clearings (25 to 50ha) where elephants congregate for multiple purposes. The recording devices sample audio signals at a rate of 2000 (12-bit resolution) or 4000Hz (16-bit) and can detect elephant calls up to approximately 0.8 km away. As is typical in bioacoustic applications the animals are detected infrequently, and different locations have variable density, see Table \ref{tab:data_stats}. Additionally, multiple other sources of sound are recorded, both man-made and natural. For example, Ceb4 is close to a road and the recordings include signals associated with logging and gunshots.

\subsection{Acoustic Characteristics of the Dataset}

\begin{figure}
\centering
\includegraphics[width=1.0\columnwidth]{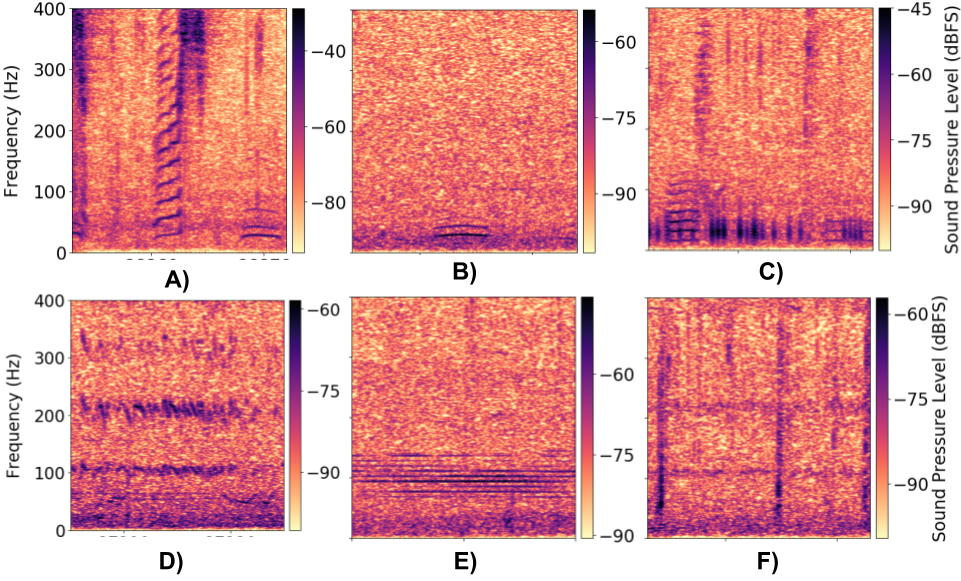}
\caption{Examples of the diversity of acoustic signals encountered in sound streams from Central African forest environments. \textbf{A)} an elephant call combining both tonal and chaotic (broadband) sounds, often produced in agonistic situations. \textbf{B)} an elephant rumble with few harmonics (source far from microphone and/or low amplitude). \textbf{C)} signals emitted by a dwarf crocodile (Osteolaemus tetraspis), including some harmonics similar to those of elephants. \textbf{D)} the buzzing of insects \textbf{E)} a motorized vehicle \textbf{F)} sound of splashing of water as elephants move through a stream.}
\label{fig:variability}
\end{figure}

The primary mode of communication among elephants is a low-frequency vocalization known as a rumble, typically lasting between 2 and 8 seconds. These sounds have distinct frequency characteristics, with a low fundamental frequency (8 - 34Hz), often several higher harmonics, and slight frequency modulation. A typical recording is shown in Figure \ref{fig:typical_calls}. At large gatherings, multiple elephants often make simultaneous or overlapping calls (see for example Figure \ref{fig:typical_calls} where two calls overlap). Other complications are the variability of the dataset, for example, some recording sites are close to logging concessions which are often visited by motorized vehicles which become recorded. Natural sources of noise include heavy wind, rainfall, insects chirping and thunderstorms, see Figure \ref{fig:variability} for further examples.

\subsection{Labeling} \label{labeling}

The labeling of rumbles to be used in the training and testing of detection algorithms was done by both experts and trained volunteers at the Elephant Listening Project. The volunteers were recruited by a combination of work-study positions and information spread via word-of-mouth and were asked to identify individual elephant calls and their temporal extent in the recording. Positive labeling was based on a set of criteria developed by experts with more than ten years of experience with forest elephant vocalizations and experience with potentially confusing environmental sounds. Volunteers followed a detailed training program that concluded with them labeling rumbles in two 24 hour long test sound files. The labels generated by the volunteers for the test files were compared to those of an expert. If the results were within 5\% of each other, the volunteer was considered trained; if not, he/she repeated the process on other sound files until the 5\% or less difference was achieved. The occasional further review of volunteer labeling efforts by the experts maintained reasonable consistency among all labelers (reliability $> 98\%$). Statistics about the dataset and the labeling can be seen in Table \ref{tab:data_stats}. To facilitate online crowdsourcing, we have created an online labeling application for labeling. The website contains a tutorial where participants can first learn about the characteristics and variations of elephant calls and other sounds that might occur in recordings. The tutorial is publicly available at \url{www.udiscover.it/applications/elp/tutorial.php}. Once trained, participants can then label elephant calls in audio segments by using the application's annotation tool, see Figure \ref{fig:annotation_tool} in the Appendix. By giving the same spectrum to multiple participants one can gauge the accuracy of individual users and can encourage truthful responses. These issues will be addressed further in future work.

\begin{table}
\begin{center}
    \begin{tabular}{ l | l l l l l }
    \hline
    Location & \thead{Dates \\ Collected} & \thead{Labelled \\ hours} & \thead{Num. \\ calls} & \thead{Apx. \% \\ Calls} \\ \hline
    Ceb1 & 09/04 - 11/06 & 1870 & 52810 & 0.784 \% \\
    Ceb4 & 08/06 - 11/03 & 1280 & 23038 & 0.500 \% \\
    Jobo & 09/05 - 11/06& 1437 & 28609 & 0.553 \% \\
    Dzan & 11/04 - 12/02 & 312 & 63792 & 21.8 \% \\ \hline
    \end{tabular}
\end{center}
\caption{The statistics of the datasets by location. The Apx. percentage of calls refer to what portion of the audio recordings contained elephant calls. The dates are given in YY/MM format. }
\label{tab:data_stats}
\end{table}

\section{Classification and Segmentation} \label{sec:exp}

The simplest and most straightforward problem for passive acoustic monitoring is that of detection. Given a short audio-clip we want to classify it as containing a signal produced by the species of interest (in this case an elephant rumble) or not. This setting has been considered by many previous authors \cite{mac2018bat,nichols2016marine,bittle2013review}, and is a crucial stepping stone toward using PAM for population surveying and monitoring. A similar setting we consider is one of segmentation where we want to classify each discrete time step as belonging to an elephant call or not.

\subsection{Data Processing}

Given a specific location where a recording device is placed, say "Ceb1" in Table \ref{tab:data_stats}, we extract all unique elephant rumbles recorded. The calls will in some cases overlap and if so we consider them as two or more unique calls. To facilitate a homogeneous dataset, we extract signals of the fixed length 25.5 sec and remove the handful of calls that are longer than this. We then extract empty regions of the same length, that does not overlap with any elephant calls, uniformly at random, we extract as many empty regions as there are elephant calls. The combined dataset of calls and empty frames is then split uniformly and randomly into a testing and training set, where additional augmentation might be performed on the training set. Given these fixed sized windows of audio recording, we transform them into the frequency domain via FFT. Using the signals down-sampled to 1000 Hz, we use a window size of 512 and hop-length of 384 we use FFT to transform the audio signal into the frequency domain. All frequency bands above 100Hz are removed, as the set of recorded elephant calls rarely have significant signals above such frequency because of signal attenuation. This gives us a 64 time-steps by 47 frequency bands tensor, and we specifically choose time-steps to have the shape be a power of two which can be beneficial for training on GPUs. One has to be slightly careful when normalizing the dataset as the signals are very sparse, we have found that subtracting the mean of all frames containing no calls and then dividing the signal by the median call intensity works well.

\begin{table}
\begin{center}
    \begin{tabular}{ l |  l  l  l  l  l }
    \hline
    Location & SVM & RF & ADA-grad & DNN (\cite{mac2018bat}) & Densenet + rnd crop \\ \hline
    Ceb1 & 77.22 & 76.73 &  77.01 & 91.11 & 93.40 \\
    Ceb4 & 70.21 & 69.82 &  71.50 & 90.15 & 93.68 \\
    Jobo & 76.92 & 76.49 &  76.86 & 91.67 & 94.30 \\
    Dzan & 72.21 & 69.79 &  70.86 & 75.86 & 77.51 \\ \hline
    Avg. & 74.14 & 73.21 &  74.06 & 87.20 & 89.72 \\ \hline
    \end{tabular}
\end{center}
\caption{The classification accuracy on the test-set for different algorithms at different locations.}
\label{tab:accs}
\end{table}

\subsection{Neural Network Architectures and Training} \label{sec:training_params}

Audio-clips are approximately time-invariant, i.e., an elephant call will sound the same no matter if it starts after 1s or 3s. This approximate symmetry suggests the use of convolutions over the time dimension would be successful, we have however found it beneficial to additionally perform convolutions over the frequency dimension. This corresponds to an approximate pitch-invariance, meaning that elephant calls from different elephants sound similar except for a uniform pitch change. Given this two-way convolution, we adopt the state-of-the-art network architecture Densenet \cite{huang2017densely}, which is a standard convolutional network with skip connections between all layers. For training this architecture we use best practices from neural network training \cite{lecun2012efficient}, we consider SGD with momentum and weight decay, iteratively lowering the learning rate as performance plateaus and using cross-entropy as the loss. We further introduce data augmentation by adapting the technique of randomized cropping, which is typically used on image classification, to sound classification. The 64 time-steps are padded by 8 on both sides, and we feed a random 64 step long subsequence. Exact parameters are given in table \ref{tab:hyperparams} in the Appendix. For our segmentation setting we train a standard long short-term memory (LSTM) network \cite{lstm}, additionally, we use the same LSTM network with an initial one-dimensional convolutional layer on the frequency dimension. This convolution layers uses 25 filters and thus outputs a feature vector of length 25 for each time step, which is then fed into the LSTM network. For training this architecture we use the ADAM optimizer and cross-entropy as the loss.

\begin{figure}
\centering
\includegraphics[width=1.0\columnwidth]{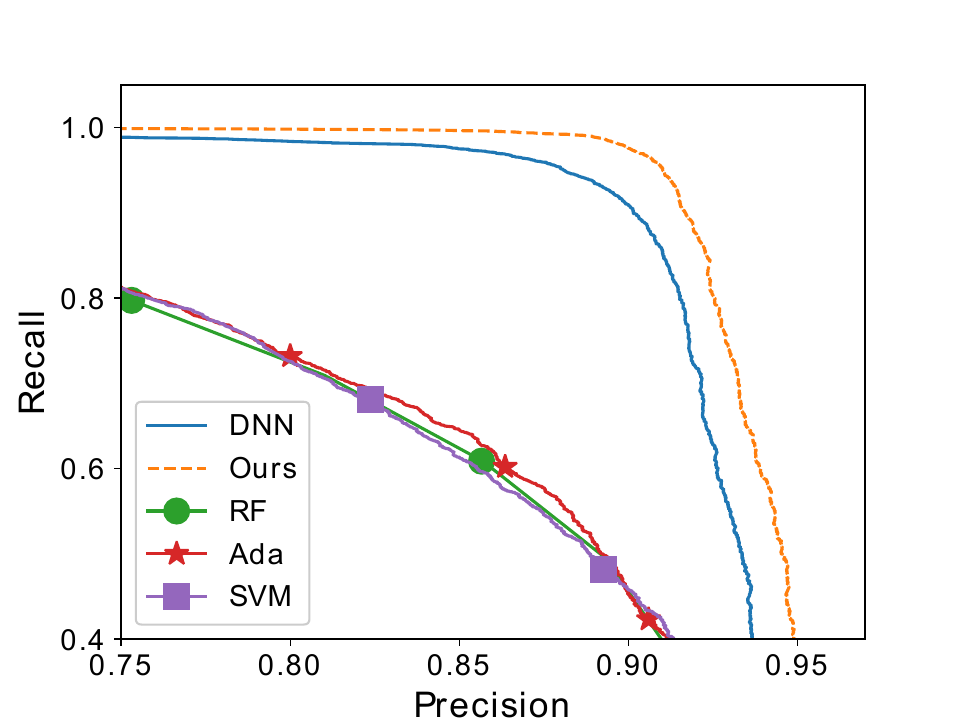}
\caption{We here illustrated the precision-recall curve of the various classifiers we consider here. All algorithms based upon the MFCC features perform relatively poorly, whereas the classic neural networks and Densenet specifically achieve much higher scores.}
\label{fig:pr-curves}
\end{figure}

\subsection{Results}

We compare our approach to previously proposed methods in bioacoustic monitoring, both modern and classical. From the latter category, we consider the method of \citeauthor{dufour2014clusterized}, where rich features based upon the MFCC coefficients are extracted and then fed into an SVM \cite{hearst1998support}. The MFCC coefficients are similar to the FFT, where the signal is decomposed into frequency parts, however, the filters and masks used are much more sophisticated. Given these coefficients for each time step, the features for the entire audio clip are the mean, variance and derivative of the coefficients for each time step across the entire audio clip. Additionally, we consider feeding these features into a Random Forest classifier \cite{liaw2002classification} and an ADAGrad classifier, essentially emulating the approach of \citeauthor{ross2014random} although with slightly different features. The second type of baselines we consider are convolutional neural networks, where we use the architecture (and training parameters/schedule) of \citeauthor{mac2018bat} proposed for classifying the vocalizations of bats. The architecture is classical, meaning no skip-connections are used and dropout is not used \cite{srivastava2014dropout} for regularization, see \citeauthor{mac2018bat} for details. Results can be viewed in Table \ref{tab:accs} where we see that our methods consistently outperform baselines across every location. A more nuanced picture over the precision and recall is given in Figure \ref{fig:pr-curves}. For the segmentation setting, we compare our convolution-LSTM hybrid network's performance to that of only an LSTM network and show how the hybrid methods perform much better in terms of accuracy, see Table \ref{tab:accRNN}.

\begin{table}
\begin{center}
    \begin{tabular}{ l | l  l }
    \hline
    Location & LSTM & conv-LSTM \\ \hline
    Ceb1 & 70.50 & 95.24 \\
    Ceb4 & 70.43 & 90.54 \\
    Jobo & 67.49 & 92.12 \\
    Dzan & 70.54 & 88.95 \\ \hline
    Avg. & 69.74 & 91.71 \\ \hline
    \end{tabular}
\end{center}
\caption{The classification accuracy on the test-set for the segmentation task, given for different algorithms at different locations.}
\label{tab:accRNN}
\end{table}

\section{Compression}

\begin{figure*}
\centering
\includegraphics[width=1.9\columnwidth]{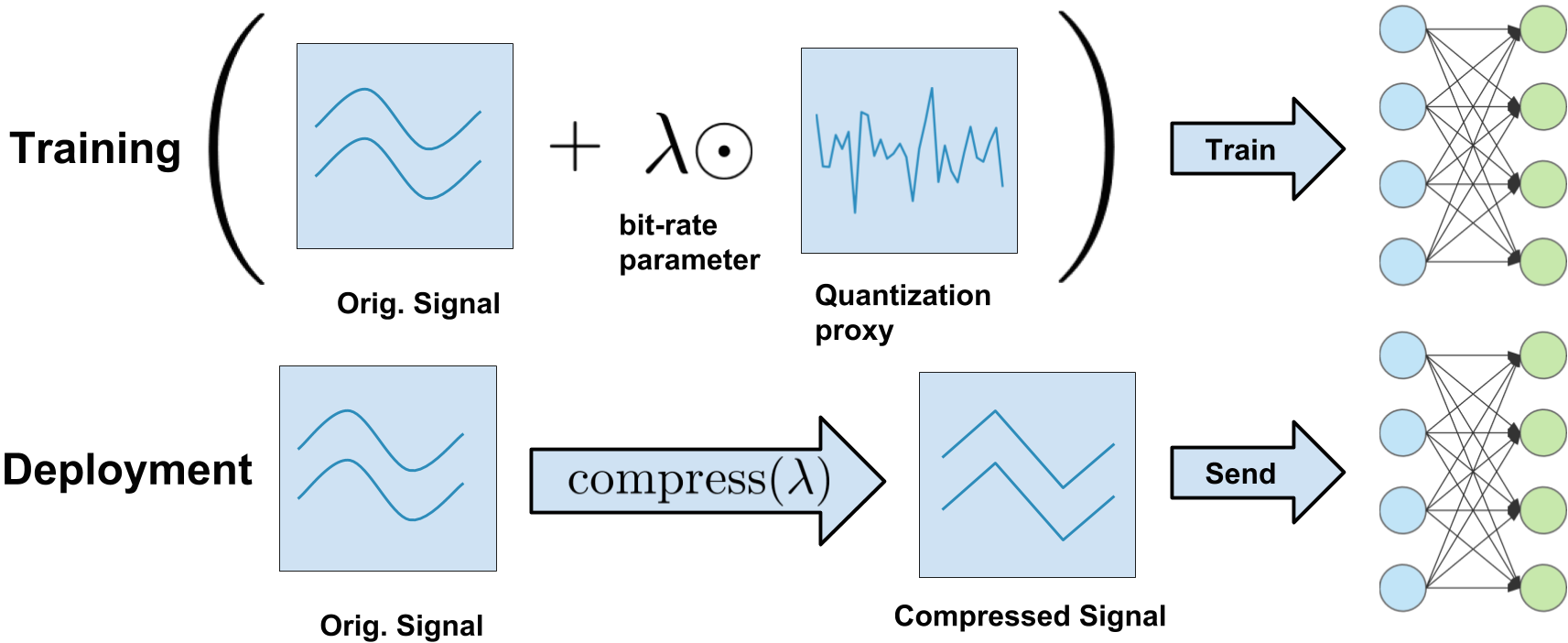}
\caption{The main idea behind our end-to-end compression scheme is to introduce a continuous bit-rate vector $\lambda$ and i.i.d. noise that serves as a proxy for the quantization error. By optimizing $\lambda$ one can adjust the quantization level for different frequency bands, which can be optimized jointly with a neural-network classifier to find compression strategies that result in signals that are useful for classification. At deployment, the bit-rates of individual frequency channels are used for compression at the recording devices so that data transfer is minimized.}
\label{fig:compr_scheme}
\end{figure*}

\subsection{Background}

The ultimate aim of passive acoustic monitoring is to provide accurate real-time detections of elephant vocalizations and threats. It is infeasible to perform neural network computations on the recording devices, and hence the devices need to send their data over the wireless networks of sub-Saharan Africa. Unfortunately, wireless infrastructure is largely relatively poor or absent in this area of the world \cite{aker2010mobile}, available bandwidth is small and data-transfer is expensive. To make real-time passive acoustic monitoring cost efficient, one has to transfer only the most relevant information across the wireless network.

A natural strategy for reducing the data-transfers across the wireless network is to compress the acoustic data. Most lossy compression codecs crucially rely on the specifics of the human auditory system to remove data that are irrelevant to the experience of a human listener. For example, it is well known that the sensitivity of the human auditory system varies with frequency \cite{painter2000perceptual}, and hence many lossy compression algorithms remove low-frequency components or simply use a low bit-rate for them. In the context of elephant monitoring, this is a poor strategy since the elephants communicate by low-frequency rumbles. It is clear that we need to develop compression strategies uniquely suited for the elephant calls and for the neural networks that will analyze them. As neural networks are well known to be resistant to minor random perturbations \cite{micikevicius2017mixed} lossy compression is a promising avenue. Additionally, as passive acoustic monitoring has applications to many species, from small birds \cite{bardeli2010detecting} to marine mammals \cite{bittle2013review}, data-driven approaches such as ours avoids the laborious process of manually crafting audio codecs and can easily be adapted to new species. It does not require any hand-crafted features or any specific information regarding the structure of animal vocalization (save for an approximate frequency range, information that is easily obtainable for most species), and one would “only" need training data to adapt our framework to other species.

\begin{figure}
\centering
\includegraphics[width=1.0\columnwidth]{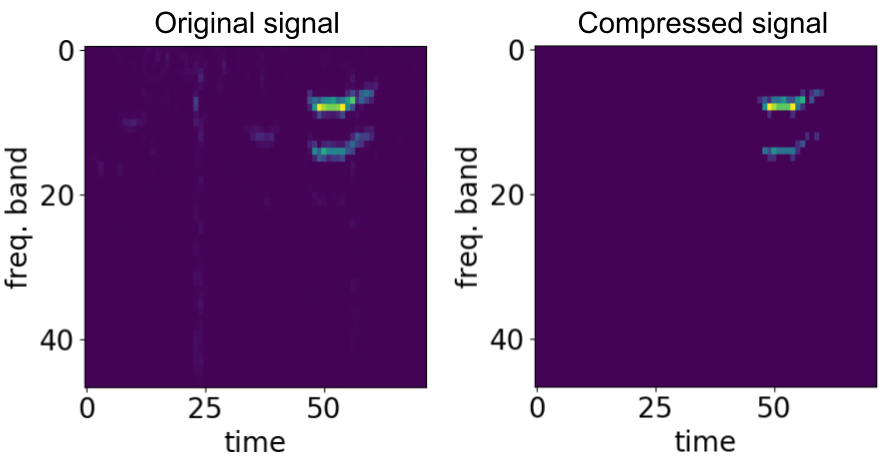}
\caption{We here illustrate an example of quantization of a signal with elephant calls with extremely low bit-rate. Background signal almost disappears with quantization while the elephant call loses much of its nuances.}
\label{fig:quant_err}
\end{figure}

\subsection{End-to-end differentiable compression codecs}

As opposed to typical audio compression applications, the listener in our setup is not a human, additionally, the frequency spectrum is vastly different. To study this phenomenon in isolation and achieve a simple setup we only consider compression in terms of the different frequency bands. Other aspects of lossy compression, for example, lossless compression on top of lossy strategies, can be added to all methods we consider. We assume that the one-dimensional $X$ that describes the sound waves has been transformed via FFT as a pre-processing step into $\hat{X}$, and consider the problem of assigning bit-rates to the different frequency bands. Simple operations such as FFT and bit-truncation can easily be implemented on the rudimentary hardware of the recording devices. We propose a method that jointly optimizes for low bit-rates of the frequency channels and high classification accuracy.

Our algorithmic setup is illustrated in Figure \ref{fig:compr_scheme}. We want to assign different bit-rates to different frequency-channels, which we achieve by simply truncating the bit representation of elements of the channels, which lowers the precision. Our key insight is to exchange a non-differentiable bit-truncation by a differentiable proxy -- we simply model truncation as additive Gaussian noise, a common model of quantization error \cite{gray1998quantization}. We let the components of the vector $\lambda$ denote the bit-rates of various frequency channels, and let $\beta$ be a matrix with dimensions ${t \times f}$ with independent standard Gaussian entries, where there are $t$ time-steps and $f$ frequency bands. The truncation error is the proportional to by the matrix $\exp(-\lambda) \odot \beta$, where the entries $(i,j)$ are equal to $\exp( -\lambda_j)\beta_{ij}$. This ensures that the additive errors in the original elephant spectrogram $\hat{X}$, which models bit-truncation, are independent but that each frequency band has its own error scale. The input to the neural networks is thus $\hat{X} +  \exp(- \lambda ) \odot \beta$, and we simultaneously optimize the network parameters $\omega$ for large classification accuracy and the total bit-rate which is simply expressed as $\sum_i \lambda_i$, balancing these two objectives with the hyper-parameter $\mu$. The loss can be written as

\begin{equation}
\label{eq:opt_prob}
\underset{\substack{\beta \sim N \\ (\hat{X},y) \sim D}}{\mathbb{E}} \bigg[ L \bigg(y, \textrm{DNN}_\omega \big( \exp(- \lambda ) \odot \beta + \hat{X} \big) \bigg) \bigg] + \mu \sum_i \lambda_i
\end{equation}

Here the dataset $D$ contains tuples $(\hat{X},y)$ of data $\hat{X}$ and labels $y$, $L(y, \hat{y})$ denotes the loss function used to measure goodness of fit between ground-truth label $y$ and estimated label $\hat{y}$. The function $\textrm{DNN}_\omega$ gives the output of the trained neural network with network parameters $\omega$. We again use the cross-entropy for the loss function. This function can be optimized via SGD, where we exchange the expectation $\mathbb{E} [\; \cdot \;]$ by sample averages.

\subsection{Experiments}

We compare different compression strategies by how well they transmit the important information as measured by how well a classifier can be trained to classify compressed elephant spectrograms given a fixed bit-rate. For all compression strategies, we will use the Densenet model of earlier sections. The original Fourier signal has elements put into one of the $2^{32}$ bins represented as 32 bit signed integers, lowering the bit-rate simply corresponds to removing the least significant bits with the sign bit is removed last. This has the effect of quantizing the signal and removing small variations in signal strength while keeping the large variations (see Figure \ref{fig:quant_err}). We enforce that no less than 5 bits are used for each frequency band as the dynamic range of the audio signal has the effect of completely erasing the signal for smaller bit-rates. For assigning bit-rates via optimizing \eqref{eq:opt_prob} we use the same Densenet architecture as for evaluating the compression quality, and train it with the same parameters as in earlier sections and with $\mu = 10^{-7}$. To ensure specific total bit-rates we assign bit-rates to various frequency bands proportional to the values of the components of $\lambda$. We compare our method against the method of assigning bit-rates proportional to the sensitivity of human hearing, using the well-known model of how human auditory sensitivity vary with frequency of \cite{painter2000perceptual}. The proportional allocation excludes the 5 bits needed for the dynamic range of the signal. The results for various locations and bit-rates are given in Table \ref{tab:compr_rates}, where we can clearly see that our proposed method achieves superior performance for the same bit-rates. For very small and very large bit-rates the difference becomes smaller. Implementing our method leads to data compression of a factor roughly 116 compared to näively storing the 1000Hz signal in 32-bit floating point numbers while achieving little performance degradation. These savings are significant for the often poor wireless networks of sub-Saharan Africa.

\begin{table}[t]
\begin{center}
    \begin{tabular}{ l |  l  l  l  l }
    \hline
    Method / Bit-rate & Ceb1 & Ceb4 & Jobo & Dzan \\ \hline
    Ours / 47 & 84.57 & 83.98 &  86.31 & 78.43 \\
    Human / 47 & 83.62 & 81.31 &  85.76 & 69.44 \\ \hline
    Ours / 141 & 92.81 & 92.21 &  93.19 & 77.96 \\
    Human / 141 & 86.61 & 91.90 &  90.32 & 73.51 \\ \hline
    Ours / 235 & 93.05 & 93.11 &  93.84 & 77.46 \\
    Human / 235 & 90.25 & 92.34 &  91.64 & 76.93 \\
    \end{tabular}
\end{center}
\caption{The classification accuracy on the test-set for the given bit-rates at various locations.}
\label{tab:compr_rates}
\end{table}

\section{Related Work}

\subsection{Bioacoustics}

The field of bioacoustics has for a long time been interested in automatic approaches towards detecting and classifying animal vocalizations with the ultimate goal to accurately survey population size and behavior \cite{mcdonald1999passive}. As sound waves attenuate less in water, passive acoustic monitoring can cover vast underwater areas. Much effort has been in terms of large marine animals with characteristic vocalizations -- predominately various whale species (Humpback, right \cite{thode2017using}, Baleen \cite{baumgartner2011generalized}, Blue and Fin \cite{vsirovic2007blue}) and dolphins \cite{erbs2017automatic}. Acoustic signals are the primary mode of communication for many marine species and for large gatherings vocalizations typically overlap which together with long reverberation times becomes challenging. Techniques used to overcome these issues include blind source separation \cite{zhang2017blind}, pitch-tracking via dynamic programming \cite{baumgartner2011generalized} and kernel methods \cite{thode2017using}.

On land, efforts towards bioacoustics have primarily focused on various bird species, owing to the characteristic songs many of them use for mating and communication. As bird species typically have unique songs, PAM makes it possible to accurately survey populations of endangered species, whereas using direct visual observations becomes problematic for species that are small and/or occupy canopies \cite{bardeli2010detecting}. Popular strategies include SVMs based upon MFCC \cite{dufour2014clusterized}, segmentation via deep learning \cite{koops2015automatic} and dictionary learning \cite{salamon2017fusing}. Beyond birds, insects \cite{ganchev2007automatic}, bats \cite{mac2018bat} and monkeys \cite{turesson2016machine} have all been considered. Elephants have been studied from a similar perspective to ours by \citeauthor{geoff}. For many of these species, especially many birds, the vocalizations occupy a relatively small frequency band making models less sensitive to noise and intra-population variability in vocalizations, hence making them unsuitable for elephant monitoring.

\subsection{Machine-learning for Audio}

Machine learning for audio-signals has primarily focused on human speech due to applications such as virtual assistants, automatic transcription, and translation. For a long time, mainstream research was primarily propelled by using the EM-algorithms for training Hidden-Markov-Models \cite{hinton2012deep}. Features for audio input could often be encoded via MFCC \cite{sahidullah2012design}, and rich distributions could be represented via Gaussian-Mixture-Models \cite{juang1986maximum}.  While using neural networks for acoustic applications was conceived more than 25 years ago \cite{bourlard2012connectionist}, it was in only 2009 that deep learning approaches were shown to be competitive with more traditional ``hand-crafted" machine learning approaches \cite{mohamed2009deep}. Deep learning has now gained mainstream traction and it has become the dominant paradigm. State-of-the-art speech recognition often relies on recurrent neural networks \cite{graves2014towards} \cite{sak2014long}, where convolutional layers can automatically extract features \cite{sainath2015convolutional}. Beyond speech recognition, deep learning for acoustic sensing in smartphones has been investigated \cite{lane2015deepear}.

\subsection{Compression}

Compression for acoustic signals has been studied for a long time due to applications such as storing music on handheld devices and sending human conversations across networks, and many audio compression methods rely on essentially handcrafted features, for example, wavelets \cite{jagadeesh2014psychoacoustic}. Most methods for lossy compression of audio has the goal of ensuring signals are audible to humans, and hence most models are based upon the models of human hearing, so-called psychoacoustic models. A salient feature of human hearing is that its sensitivity varies with frequency \cite{painter2000perceptual}, a common strategy is to transform the audio-signal with the modified discrete cosine transform (MDCT) and address frequency bands individually. Another phenomenon of human hearing is called simultaneous masking where signal $A$ can make signal $B$ (which is of a different frequency and intensity) inaudible \cite{jagadeesh2014psychoacoustic}. 

While traditional compression schemes have typically relied on handcrafted features, the advent of deep learning has spurred interest in data-driven approaches to compression. Previous research has primarily focused on images and video, proposing various continuous and differentiable proxies for entropy and quantization, see for example \cite{balle2016end} and \cite{agustsson2017soft}. The only work on audio compression known to the authors is on human speech \cite{kankanahalli2017end} which has is different in terms of frequency distribution, complexity, and dataset cleanliness; the proposed architecture relies on softmax quantization.

\section{Future Work and Conclusions}

Managers of protected areas designed for the forest elephants are interested in better conservation tools but need to see definitive proof of their efficacy. If useful information about elephant populations and human encroachments can reach managers within a reasonable timeframe, the potential to expand acoustic monitoring across the Congo Basin becomes a reality. Collaboration with managers is thus instrumental in developing a rapid work-flow for the current acoustic monitoring project in northern Congo, which covers 1500 square km of rainforest and generates seven terabytes of sound data quarterly. We hope that these proof-of-concept demonstrations of how various AI techniques can inspire future work on PAM, with the ultimate goal of real-world implementation.

In this work, we have introduced a dataset of elephant calls recorded in the wild aimed at promoting interest and progress in automatic methods for passive acoustic monitoring and discussed our methods for labeling it. Using modern neural network architecture, state-of-the-art training regimes and data-augmentation techniques we have shown how to improve upon previously proposed method for passive acoustic monitoring. Additionally, we have addressed how wireless network infrastructure is often lacking in sub-Saharan Africa data transfer quickly becomes a bottleneck for real-time systems. To circumvent this issue, we have introduced a novel scheme for jointly optimizing bit-rates and prediction accuracy, which beats a baseline based upon models of human hearing.

\subsection{Acknowledgements}
We would like to thanks the Elephant project, the Cornell Lab of Ornithology its volunteers and the Wildlife Conservation Society. This work is supported by NSF Expedition CCF-1522054 and ARO DURIP W911NF-17-1-0187. The work of PHW was supported by the U.S. Fish and Wildlife Service and the Robert G. and Jane V. Engel Foundation.

\appendix

\section{Appendix}

\begin{table}[h]
\begin{center}
    \begin{tabular}{ l | l }
    \hline
    Parameter & Value \\ \hline
    init. learning rate & 0.1 \\ 
    SGD momentum & 0.9 \\ 
    batch size & 64 \\ 
    initialization & kaiming \\ 
    weight decay & 0.0001 \\ 
    loss function & cross-entropy \\ \hline
    \end{tabular}
\end{center}
\caption{Hyper-parameters used for training.}
\label{tab:hyperparams}
\end{table}

\begin{figure}[h]
\centering
\includegraphics[width=.95\columnwidth]{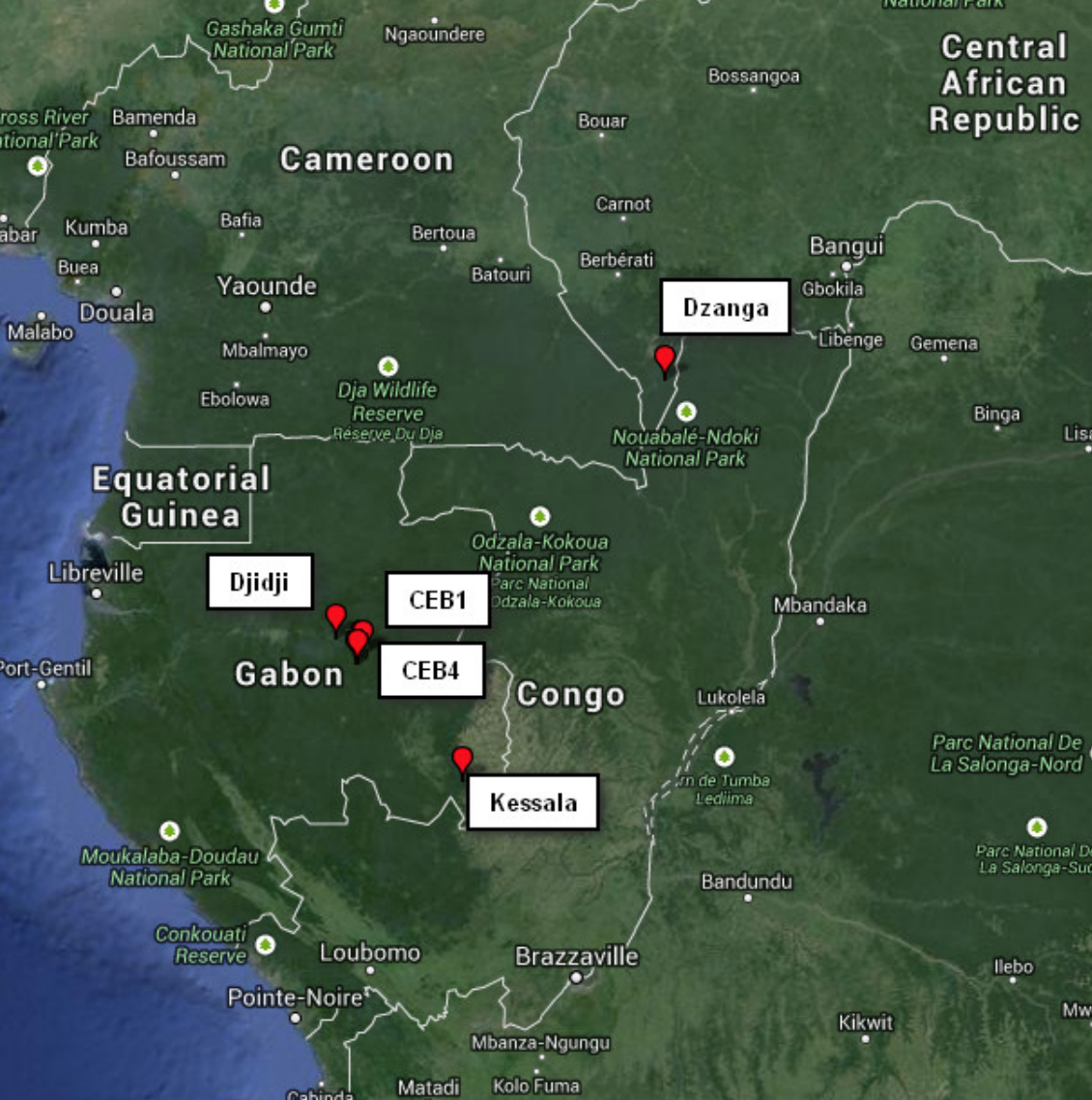}
\caption{The sites from which the elephant recordings were collected, as seen via Google Earth. All areas were close to forest clearings, additionally, one is close to a river while CEB1 and CEB4 are within logging concessions. Note that this paper did not use sounds from all sites in this map.}
\label{fig:map}
\end{figure}

\begin{figure}[h]
\centering
\includegraphics[width=.95\columnwidth]{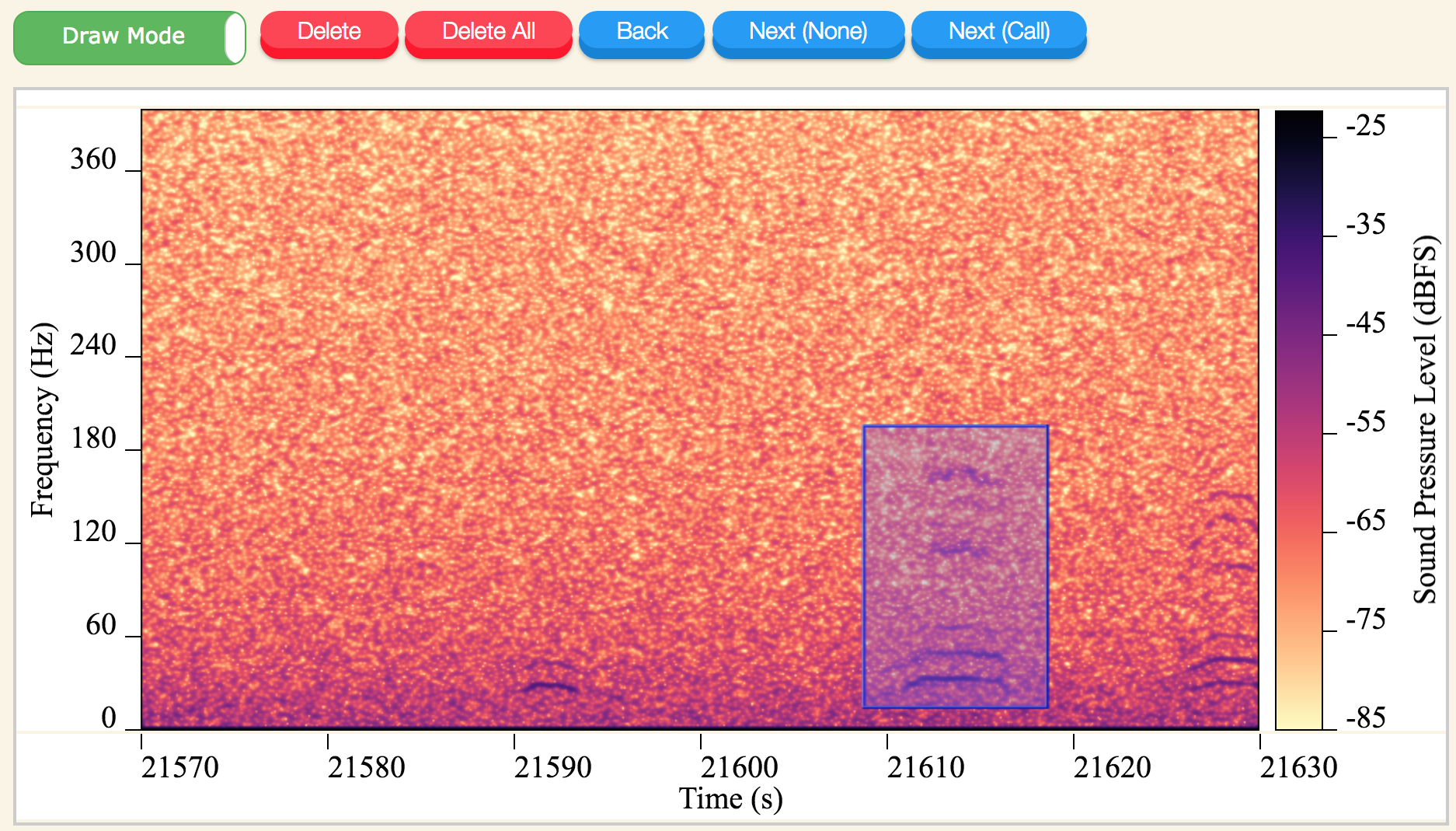}
\caption{We here show the interface of online annotation tool, which enables crowdsourcing of labeling efforts.}
\label{fig:annotation_tool}
\end{figure}

\fontsize{9.0pt}{10.0pt}
\selectfont
\bibliography{eleph}
\bibliographystyle{aaai}
\end{document}